\documentclass[12pt]{article}

\RequirePackage[OT1]{fontenc}
\usepackage{amsthm,amsmath,natbib}
\usepackage{array}
\usepackage{float}
\usepackage{booktabs}
\usepackage{tabularx}
\usepackage{xcolor}
\usepackage{color}
\RequirePackage[colorlinks,citecolor=blue,urlcolor=blue]{hyperref}
\usepackage{alltt}
\usepackage{fancyvrb}
\usepackage{algorithm2e}
\usepackage{pdfpages}
\usepackage{multirow}
\usepackage{lineno}
\usepackage{comment}
\usepackage{verbatim}
\usepackage{lscape}
\usepackage{graphicx}
\usepackage{eurosym}
\usepackage{gensymb}
\usepackage{graphicx}

\def\bi{\begin{itemize}}
\def\ei{\end{itemize}}
\def\be{\begin{equation}}
\def\ee{\end{equation}}

\begin{document}
\thispagestyle{empty}
\baselineskip=27pt
\vskip 4mm
\begin{center} 

{\Large{\bf From scenario-based seismic hazard to scenario-based landslide hazard: fast-forwarding to the future via statistical simulations}}

\end{center}

\baselineskip=12pt
\vskip 3mm

\begin{center}
\large
Luigi Lombardo$^{1*}$, Hakan Tanyas$^{2,3}$
\end{center} 

\footnotetext[1]{
\baselineskip=10pt University of Twente, Faculty of Geo-Information Science and Earth Observation (ITC), PO Box 217, Enschede, AE 7500, Netherlands}

\footnotetext[2]{
\baselineskip=10pt Hydrological Sciences Laboratory, NASA Goddard Space Flight Center, Greenbelt, MD, United States}

\footnotetext[3]{
\baselineskip=10pt USRA, Universities Space Research Association, Columbia, MD, United States}

\baselineskip=16pt

\begin{center}
{\large{\bf Abstract}}
\end{center}

Ground motion scenarios exists for most of the seismically active areas around the globe. They essentially correspond to shaking level maps at given earthquake return times which are used as reference for the likely areas under threat from future ground displacements. 
Being landslides in seismically actively regions closely controlled by the ground motion, one would expect that landslide susceptibility maps should change as the ground motion patterns change in space and time. However, so far, statistically-based landslide susceptibility assessments have primarily been used as time-invariant.In other words, the vast majority of the statistical models does not include the temporal effect of the main trigger in future landslide scenarios. 
In this work, we present an approach aimed at filling this gap, bridging current practices in the seismological community to those in the geomorphological and statistical ones. More specifically, we select an earthquake-induced landslide inventory corresponding to the 1994 Northridge earthquake and build a Bayesian Generalized Additive Model of the binomial family, featuring common morphometric and thematic covariates as well as the Peak Ground Acceleration generated by the Northridge earthquake. Once each model component has been estimated, we have run 1000 simulations for each of the 217 possible ground motion scenarios for the study area. 

From each batch of 1000 simulations, we have estimated the mean and 95\% Credible Interval to represent the mean susceptibility pattern under a specific earthquake scenario, together with its uncertainty level. Because each earthquake scenario has a specific return time, our simulations allow to incorporate the temporal dimension into any susceptibility model, therefore driving the results toward the definition of landslide hazard. 

Ultimately, we also share our results in a shapefile where we report the mean (and uncertainty) susceptibility of each 1000 simulation batch for each of the 217 scenarios.      

\baselineskip=10pt

\par\vfill\noindent

\newpage
\baselineskip=16pt

\section{Introduction}

The definition of landslide hazard originally proposed by \citet{varnes1984} and later updated by \citet{guzzetti1999landslidehazard} features a spatial component aimed at defining ``where'' slope failures can be expected, a temporal component, aimed at defining ``when'' or how frequently the slope instability process is expected and an additional component aimed at defining ``how large or destructive'' a single landslide or a landslide population may be.

The susceptibility is by far the most studied element among the three \citep{reichenbach2018}, whereas the ``size or destructiveness'' is the least common although it has been highlighted to be crucial even in international guidelines \citep{fell2008guidelines}.  

The temporal component is usually investigated separately from the susceptibility, this being the case of rainfall-threshold studies \citep[e.g.,][]{aleotti2004,guzzetti2004,wu2015} or Early Warning Systems \citep[e.g.,][]{devoli2015,greco2013,guzzetti2019}. 
Because of this, few examples exist where the spatial and temporal components of the landslide hazard are combined together \citep[e.g.,][]{ghosh2012}. The typical statistically-based example corresponds to \citet{cama2015predicting} where the authors tried to calibrate a susceptibility model over past landslides and validate it over a subsequent landslide inventory. However, this approach assumes the susceptibility to be stable across time and neglects any dependence between past and future slope failures. A step forward in this direction has been proposed by \citet{samia-LandslidesFollow-2017} where the authors accounted for landslide interactions across time by simplistically building multi-temporal susceptibility models where the next model features as covariate the presence-absence signal of the previous event. However, this approach retrieves a single parameter for the whole study area to describe the temporal dependence, which is also determined in a discrete series of temporal values. The same concept has been recently brought even further by \citet{lombardo2019space} where the authors assessed the space-time landslide intensity (number of landslide per mapping unit) by featuring a temporal dependence at the Slope Unit level via a Autoregressive model. However, even in these cases, the residual space-time dependence due to the trigger is not directly incorporated in the analyses. 

Only few examples directly include the trigger effect in the susceptibility/hazard models. For instance, \citet{Nowicki2014,nowicki2018global} demonstrate how the US Geological Survey provides a statistically-based near real-time prediction by using a model that has initially estimated the effect of ground motion over the available set of global earthquake-induced landslides (EQIL) \citep{Schmitt2017,Tanyas2017}. Then by assuming that the effect is constant for other earthquakes, they plug-in the ground motion of near real-time earthquakes to produce case-specific prediction maps. 

Apart from statistical approaches, there is a large literature aiming at assessing landslide hazard via probabilistic assessments \citep[e.g.,][]{jibson2000method,romeo2000seismically,gaudio2003approach,gaudio2004time,rathje2008probabilistic,saygili2009probabilistically}.  \citet{jibson2000method} developed one of a pioneer probabilistic seismic landslide hazard assessment method to conduct a regional-scale seismic slope stability analysis. In fact, the 1994 Northridge earthquake made this progress possible as this is the first earthquake for which extensive data on engineering properties of geologic units, ground shaking, and triggered landslides were available \citep{jibson1998method,jibson2000method}. \citet{jibson2000method} used these data sets and combined shear-strength data for each geologic unit in the area with slope derived from a 10-m digital elevation model (DEM) to predict the threshold ground-shaking acceleration required for the initiation of the sliding (referred to as the critical or yield acceleration). They then predicted the resulting displacement using an empirical displacement model based on Newmark’s sliding-block method \citep{newmark1965effects} that used shaking levels recorded during the earthquake. Finally, they compared the predicted displacements to the EQIL inventory and showed that increasing predicted Newmark displacement does, in fact, correlate with increasing landslide frequency. The study provides a simple mathematical relation between predicted Newmark displacement and the probability of landsliding. This study thus provides a basic quantitative framework for using a physical modeling approach to estimate seismic landslide hazards at regional scale. In fact, this approach was also improved by adopting some of the inputs of these physical-based models from scenario-based earthquakes \citep[e.g.,][]{rathje2008probabilistic,saygili2009probabilistically}). As with most physically based methods, this simplified mechanistic approach has the advantage of more accurately reflecting the underlying processes, despite the uncertainties caused by those simplifications \citep{allstadt2017integrating}. But the geotechnical and seismic data required to apply this model are not available everywhere and can be difficult to estimate.

A similar framework is also adopted in the context of rainfall-induced landslide. \citet{kirschbaum2018satellite} demonstrate how the NASA provides near real-time prediction by using a model that has initially estimated the precipitation effect over the available set of global rainfall-induced landslides \citep{kirschbaum2010,kirschbaum2015}. Then the near real-time landslide prediction is obtained by plugging in meteorological estimates of the incoming storms.   

These methods are extremely useful if not fundamental to support the initial stages of any disaster, helping authorities to prioritize certain regions or actions as a function of the expected hazard. However, they cannot support long term planning because of their near real-time nature.  

A comprehensive landslide susceptibility or hazard assessment, helpful to plan strategies for the expected disasters in the future, should integrate both morphological information as well as the expected scenarios of the given trigger. Some examples that fit in this description can be found in \citet{ko2018,melchiorre2012} where the authors integrate to a baseline prediction model possible rainfall scenarios or, in \citet{lari2014} where the authors simulate rockfalls of different sizes. However, despite some specific cases mostly dealing with rainfall induced landslides \citep{kirschbaum2009}, scenario-based landslide hazard realizations are not as common as in other geoscience branches. For instance, in engineering or seismology it is common practice to produce scenario-based seismic hazard maps where the scenario at hand corresponds to the exceedance probability of an extreme ground motion occurring within a certain return time \citep[e.g.,][]{abrahamson2005,hanks2005,lee2000,montilla2003}.        

In this work, we took inspiration from the scenario-based studies mentioned above in relation to rainfall-induced landslides and from the scenario-based context in seismology. More specifically, we include scenario-based realizations of ground motion patterns into Slope-Unit-based landslide susceptibility models, via statistical simulation. We do this by building a Generalized Additive Model \citep[GAM,][]{goetz2015,Lombardo.etal:2018} of a binomial family using as target variable the presence/absence distribution of EQIL associated with the 1994 Northridge earthquake \citep{harp1995inventory,harp1996landslides}. We correlate this inventory to geomorphological parameters together with the Peak Ground Acceleration (PGA) of the Northridge seismic trigger \citep{worden2016shakemap}. From this reference model, we then simulate 1000 statistical realizations of the estimated susceptibility for each of the 217 ground motion scenarios in the Northridge area available at \citep{USGS2017}. As a result, we are able to draw information already developed from the seismological community, and feature it into susceptibility models that can therefore reflect the estimated susceptibility at given return times of the seismic trigger. Our susceptibility then both features a native spatial characteristic together with the coexisting temporal dimension carried by the scenario-based PGAs.   

Overall, we present our study by introducing the 1994 Northridge earthquake, the associated landslide inventory Section \ref{sec:Northridge} and the available earthquake scenarios for the study area Section \ref{sec:Scenarios}. Subsequently, we describe the modeling strategy and the simulation steps Section \ref{sec:Strategy} whereas the results are shown in Section \ref{sec:Results}. Section \ref{sec:Discuss} provides our interpretation of the framework we propose and Section \ref{sec:Conclusions} lists our concluding remarks and future perspectives. 

Supplementary material contains the results of our contribution including a shapefile where we report the mean (and uncertainty) susceptibility of each 1000 simulation batch for each of the 217 scenarios.

\section{Study Area}\label{sec:StudyArea}

\subsection{Northridge earthquake and co-seismic landslides}\label{sec:Northridge}

To carry out our analyses, we focus on the 1994 Northridge earthquake, which is considered a milestone \citep{fan2019} because this earthquake led the scientific community to make  digital seismic networks widespread across the globe \citep{wald2003shakemap}. Also, the U.S. Geological Survey (USGS)  ShakeMap system was developed \citep{wald1999trinet} following the Northridge earthquake. This system provides worldwide estimates of ground-motion parameters and thus provides a fundamental information for many EQIL modeling studies, which is still commonly used across the whole geoscientific community \citep{allstadt2017integrating}.

\begin{figure}[t!] 
	\centering
	\includegraphics[width=\linewidth]{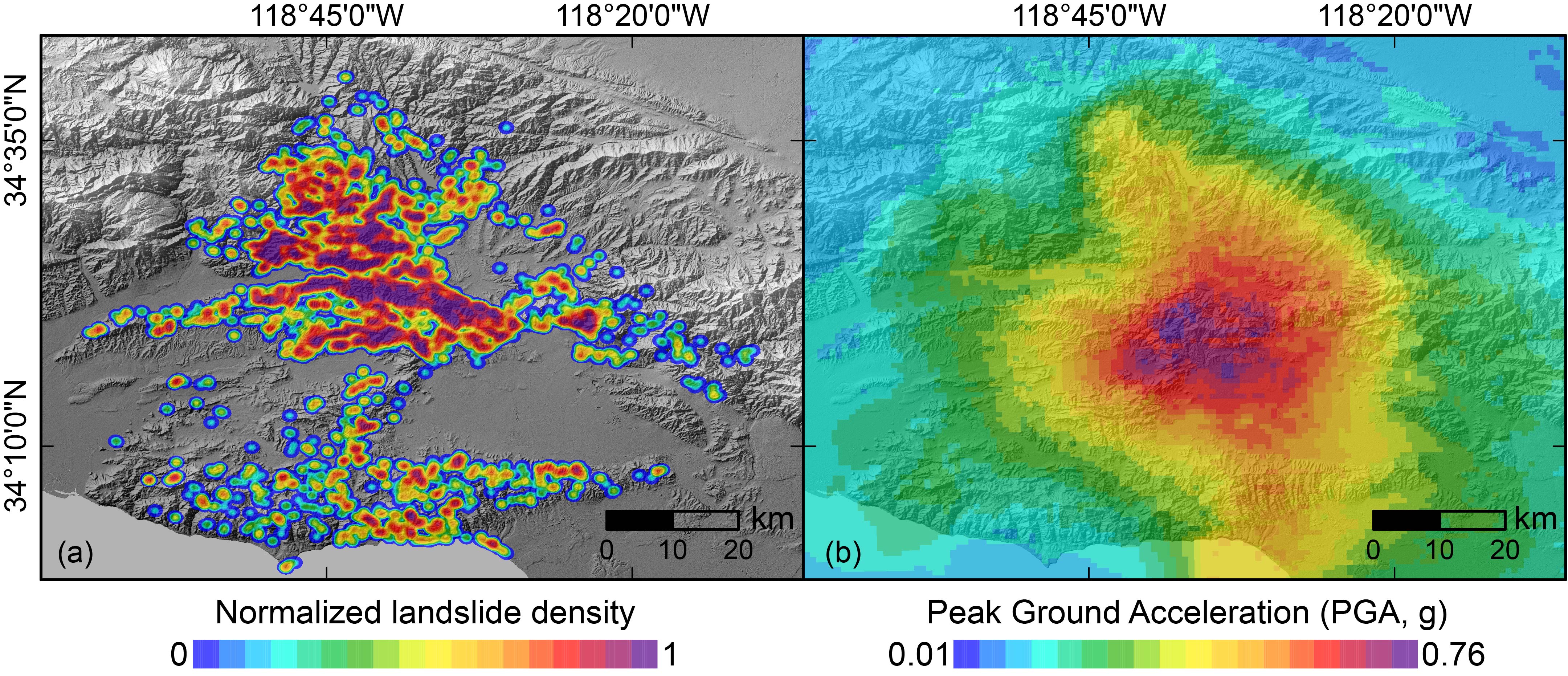}
	\caption{(a) Normalized co-seismic landslide density generated by the Northridge earthquake; (b) Ground motion due to the Northridge earthquake expressed in terms of PGA.}
	\label{Figure1}
\end{figure}

The landslides triggered by the Northridge earthquake (see Figure \ref{Figure1}a) were mapped by \citet{harp1995inventory,harp1996landslides}. They indicated that they mapped at least 90 percent of the landslide affected area using high-altitude analog aerial photography. They also reported that they most likely missed not more than about 20 percent of landslides greater than 5m and not more than 50 percent of those smaller than 5m in the maximum planar dimension. As a result, they mapped 11,111 landslides over an area of about $10,000 km^2$. The majority of the landslides were reported as shallow ($1-5 m$), highly disrupted falls and slides in weakly cemented clastic sediment \citep{harp1995inventory}. This inventory is still considered nowadays one of the most detailed EQIL datasets in landslide science because of its high quality and completeness \citep{harp2011landslide,malamud2004,Tanyas2018,TanyasLombardo2020}.  

\subsection{Scenario-based ground motion data}\label{sec:Scenarios}

The Shakemap System of the USGS not only provides actual estimates of ground motion parameters associated to earthquakes occurring all over the globe, but it also provides ShakeMaps for US based on deterministic ruptures from  U.S. National Seismic Hazard Map event catalog produced by \citet{petersen2008documentation}. 
More specifically, the available scenarios correspond to time-independent PGA maps for 2\% and 10\% probability of exceedance in 50 years for peak horizontal ground acceleration \citep{petersen2008documentation}. These scenarios originate from a subset of the possible rupture planes which are characteristic of all known active faults in a given region. 
Furthermore, a single scenario represents one single realization of a potential earthquake nucleation, assuming an underlying magnitude, hypocenter and rupture plane, although directivity is not accounted for. As a results, the USGS Shakemap System provides a comprehensive seismic hazard assessment for specific seismically active landscapes within the United States \citep{USGS2017}.  

\begin{figure}[t!] 
	\centering
	\includegraphics[width=\linewidth]{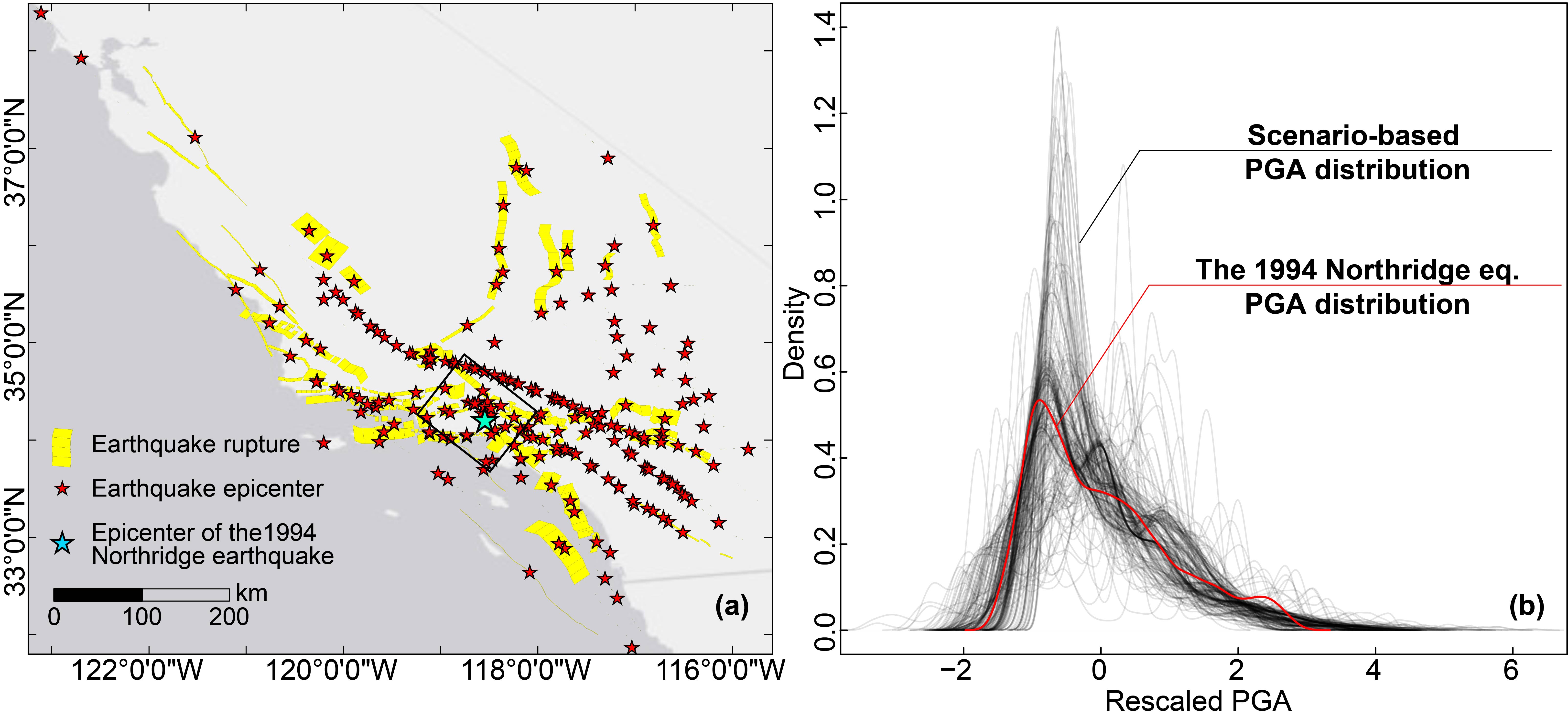}
	\caption{(a) Distribution of scenario-based earthquakes (both epicenters and rupture planes), the black polygon shows the study area for which we extracted the corresponding scenarios; (b) Probability Density Distribution of the selected and rescaled scenario-based PGAs}
	\label{Figure2}
\end{figure}

For the area affected by the Northridge earthquake, not only the specific ground motion is available (see Figure \ref{Figure1}b) but also a wide number of possible scenarios. Among the available scenarios, here we extract 217 cases which fall within the landslide affected area due to the Northridge earthquake. This information is geographically shown in Figure \ref{Figure2}a and the corresponding 218 PGA distributions are summarized in Figure \ref{Figure2}b where the PGA values have been all rescaled to emphasise similarities and differences among the possible ground motion realizations.

\section{Modeling strategy}\label{sec:Strategy}

\subsection{Mapping Unit choice and presence/absence assignment criterion}\label{sec:SU}

Any landslide susceptibility model cannot be defined without a reference mapping unit. A mapping unit correspond to the geographical object upon which any study area is partitioned. Apart from the common grid-cell choice \citep{reichenbach2018}, Slope Units (SUs) subdivide the terrain between streamlines and ridges under the constrain of Slope and Aspect within-unit homogeneity \citep{alvioli2016automatic}. In this work, we adopt SUs as the mapping unit of our choice, the reason being their ability to reliably mimic the geomorphological response to slope instability.     

We compute SUs via the software r.slopeunits made available by \citet{alvioli2016automatic} by adopting:

\begin{itemize}
    \item Shuttle Radar Topography Mission (SRTM) Digital Elevation Model \citep[DEM, of approximately 30-m resolution][]{jarvis2008hole}.
    \item a large flow accumulation threshold ($500,000m^2$) to ensure that all the possible scale-dependent SU can be examined during the calculations.
    \item a relatively small minimum SU area ($50,000m^2$) to reliably partition the space at a small to medium scale
    \item a circular variance of 0.5.
\end{itemize}

From the resulting polygonal partition of the study area, we then assigned the landslide presence condition for SU which intersect the mapped landslide polygon of the co-seismic Northridge inventory. The absence case, is of course the complementary situation.   

\subsection{Covariates}\label{sec:Covars}

Our covariate set features both morphometric and thematic properties in addition to the Northridge ground motion expressed in term of PGA. 

The morphometric properties correspond to DEM derivatives from the SRTM DEM. Specifically, we computed: \textit{i)} Slope \citep{zevenbergen1987quantitative}, \textit{ii)} Planar \textit{iii)} and Profile Curvatures \citep{heerdegen1982quantifying}, \textit{iv)} Vector Ruggedness Measure \citep[VRM,][]{sappington2007quantifying}, \textit{v)} Local Relief \citep[LR,][]{jasiewicz2013geomorphons}. 

In addition to those, we also compute the Euclidean distance from each centroid, of a squared lattice coinciding with the DEM resolution, to the nearest river channel. And, complete our covariate set with the Soil Depth map locally estimated, digitized and made available by \citep{nrcs2010soil} together with the actual SU extent measured in $m^2$.  

Notably, mapping unit of our choice does not coincide with the covariates mentioned above. For this reason, whether the resolution was higher than the SU dimension, we computed the mean and standard deviation values of each covariate distribution within the SUs. And, for Soil Depth and Northridge PGA, we opted to solely estimate the mean value because the soil map is originated from studies at scales ranging from 1:12,000 to 1:63,360 and the resolution provided by the Shakemap System is approximately 1km. 

Prior to any modeling routine, we rescaled each covariate by mean zero unit variance (substracting the mean and dividing by the standard deviation) to ensure that the covariates would share the same unitless scale, hence to ensure their comparison once we fit out model.  

\subsection{LASSO-penalized Generalized Linear Model}\label{sec:GLM}

The most common approach to build landslide susceptibility models is a Logistic Regression \citep{reichenbach2018} formulated in a Generalized Linear Model (GLM) which assumed that landslide presences/absences are distributed over space according to the Bernoulli probability distribution \citep{akgun2011,brenning2005spatial,das2010}. 

The definition of a GLM for a binomial case can be summarised as follows:

\begin{equation}\label{eq:GLMregression}
\eta(P)=\beta_0+\beta_1 x_1+\cdots+\beta_mx_m,
\end{equation}

where $\pi$ is the probability that $Y$ takes the value $1$ (or a landslide is present at a given mapping unit), $\eta$ is the logit link and $\beta$s are the estimated regression coefficients for a covariate which is assumed to behave linearly with respect to slope instability conditions.
And, by making the logit function explicit, the probability of landslide occurrence can be obtained as follows:

\begin{equation}\label{eq:GLMprob}
P = \frac{\beta_0+\beta_1 x_1+\cdots+\beta_mx_m}{1 +\beta_0+\beta_1 x_1+\cdots+\beta_mx_m},
\end{equation}

where the numerator is the linear combination of each model component, the latter being a single or multiple regression constants or the product between the regression coefficient and the corresponding vector of a given covariate values.

The choice of covariates can span over several morphometric and thematic properties in the literature \citep{budimir-SystematicReview-2015}, although some of them may not contribute to explain the landslide distribution. In such cases, maintaining a covariate set with non-informative covariates may not only increase the dimension of the model space but also bring collinearity issues and reduce the overall predictive capacity \citep{Camilo2017}. As a result, variable selection procedures are often implemented to reduce high predictor dimensionality issues. The most common variable selection found in the geoscientific literature corresponds to the Stepwise case \citep[e.g.,][]{mathew2009,cama2017improving}, although, irrespective of being foward or backward, it has been proven to be flawed in the statistical literature \citep[see,][]{harrell2015}. However, much more robust alternative variable selection procedures exist and among them, the Least Absolute Shrinkage and Selection Operator has been proven to provide not only better performance in landslide susceptibility models but also to be able to more strictly reduce the number of covariates while simultaneously addressing collinearity issues \citep[see ][and explanation therein]{amato2019accounting}.   

The LASSO selection is controlled by a penalization or shrinkage parameter, usually reported as $\lambda$. More specifically, as $\lambda$ increases the model is forced to remove redundant information from the covariate set \citep[see ][and explanation therein]{lombardo2018presenting}.    

In this work, we adopt a LASSO-penalized GLM purely to screen out non-informative covariates and subsequently test a more complex version of a GLM. We do this by using the glmnet package \citep{friedman2009} in R where by default, for 100 $\lambda$ values the functions implements a purely random 10-fold cross-validation scheme \citep[see][and explanation therein]{hastie2014} to explore the hold-out performance variability as the penalization increases.   

\subsection{Generalized Additive Model}\label{sec:GAM}

A Generalized Additive Model or GAM \citep{brenning2008,Lombardo.etal:2018} is an extension to the linear framework available in the simpler GLM case. A GAM, in addition to being able to feature linear relations (fixed effects) between covariates and landslide instances, it provides a framework to account for all sorts of nonlinear relations (random effects). The latter can correspond to categorical cases where discrete covariate classes are modeled independently from each other or, to ordinal cases where discrete covariate classes are modeled with inter-class dependence or, to latent effects over space \citep{lombardo2019numerical} and time \citep{lombardo2019space}.  

A generic formulation for a binomial case can be summarized as follows:

\begin{equation}\label{eq:GAMregression}
\eta(P)=\beta_0+\beta_1 x_1+\cdots+\beta_mx_m + f(X_n),
\end{equation}

where $f$ here stands for a nonlinear function of a covariate $X$ with $n$ discrete classes. In this work, we chose $f$ to be a random walk of the first order, or a Spline \citep{Bakka.etal:2018,conoscenti2016exploring}, which is meant to ensure that the ordinal structure between adjacent classes of a  reclassified continuous covariates is not lost. 
In this work we adopt a Bayesian version of a GAM by using the R \citep{team2013r} package R-INLA \citep[see][and explanation therein]{Rue2009,Rue.al.2017}. We chose a Bayesian version because any model (and each of its components) falling in this category is natively characterized by a distribution of values rather than the frequentist counterpart. This will be particularly relevant in the formulation of the simulation steps described in Section \ref{sec:Simulations}.  

\subsubsection{10-Fold cross-validation}\label{sec:GAM}

The landslide literature reports several cross validation schemes. Among these, very poor examples can be found in articles where a given model is merely validated once \citep[e.g.,][]{arabameri2020} and much more scientifically sound cases where k-fold cross-validation (CV) schemes are implemented \citep[e.g.,][]{steger2020} to express the variability of model performance as the training and test subsets are iteratively sampled at random. The idea behind a k-fold CV is to statistically represent a large portion of the datatest at hand (both in terms of presence-absence cases and associated covariates) hence providing a description of mean predictive skill of a given model together with the associated uncertainty. 

In this work, we opt for a special case of a k-fold CV. Specifically, we choose to randomly select from our dataset ten random partitions each corresponding to 10\% of the total, but imposing that a given SU can be sampled only once. As a result, we do not only explore a large portion of the dataset but its entirety, by calibrating over 90\% and validating onto the complementary 10\%. Notably, by rearranging the ten 10\% subsets, a fully predicted susceptibility map can be obtained, therefore reducing the effect that same SUs, and associated characteristic may yield in the calibration stage.   

\subsection{Statistical simulations}\label{sec:Simulations}

Statistical simulations are feasible in any statistical context, being frequentist or Bayesian. However, the latter case is particularly suitable to support such operations. Once a model is fitted, each component, from the intercept to fixed and random effects, is characterized by a whole parameter distribution. Therefore, it is possible to randomly generate any number of possible realization of each parameter. And, by solving for Equation \ref{eq:GAMregression} it is possible to retrieve the full spectrum of possible susceptibility estimates \citep{ferkingstad2015}. 

In this work, we use a specific type of simulations. We initially fit a model where the presence/absence distribution corresponds to the co-seismic landslides triggered by the Northridge earthquake, and where the covariate set features come covariates together with the Northridge PGA. On the basis of the distribution of each model component, we then simulate according to the scheme mentioned above. However, we use a plug-in structure \citep{stanton2013}, where the Northridge PGA values are removed and substituted by each of the 217 possible PGA scenarios for the study area. For each of the 217 PGA scenarios, we produce 1000 simulations and store the mean resulting suscptibility together with the 95\% credible interval expressed as the difference between the 97.5 and the 2.5 percentiles of the 1000 simulations.

Ultimately, we call each mean simulated susceptibility together with the posterior susceptibility estimated for the Northridge earthquake. And we use this combined information to assess which SU exhibits a landslide-prone probability across the entirety of susceptibility realizations, together with its uncertainty.

\section{Results}\label{sec:Results}

\subsection{LASSO covariate selection}\label{sec:LASSOresults}

\begin{figure}[t!] 
	\centering
	\includegraphics[width=0.6\linewidth]{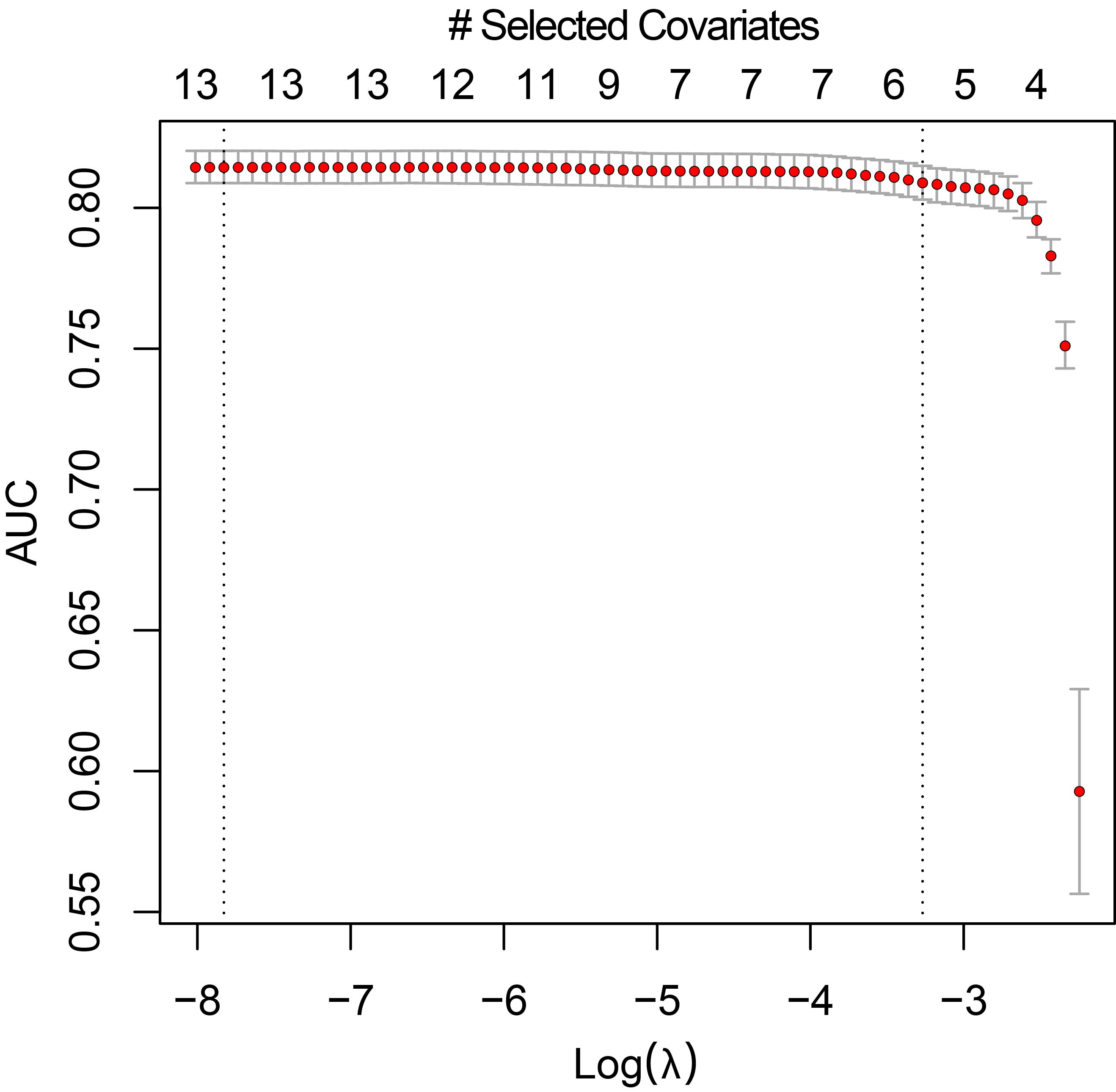}
	\caption{LASSO variable selection. The x-axis to the bottom reports the shrinkage parameter $\lambda$ in logarithmic scale (for conciseness) whereas the second axis to the top clarifies the consequent number of covariates. The y-axis shows the associated performance variability (expressed in AUC) tested in a purely random 10-fold CV. Red circles correspond to the mean AUC at specific $\lambda$ whereas the uncertainty around it is expressed as two times the standard deviation across each CV. The vertical dashed line to the left corresponds to the most conservative model. The vertical dashed line to the right corresponds to a model where the AUC has not significantly decreased compared to smaller values of $\lambda$ but larger $\lambda$ values would yield a significant drop in AUC.}
	\label{Figure3}
\end{figure}

Figure \ref{Figure3} summarizes the LASSO variable selection, where the AUC is shown to sligthly decrease as $\lambda$ increases up to a minimum number of 6 covariates (vertical dashed line to the right). Any further combination of covariates with number less than 6 is shown to rapidly weaken the model to an extent where the performance significantly drop, hence depriving the model of fundamental information to explain the presence/absence landslide distribution.   

The six selected covariates correspond to the \emph{SU Area}, \emph{VRM$_\mu$}, \emph{VRM$_\sigma$}, \emph{PGA}, \emph{Slope$_\sigma$} and \emph{Soil Depth}. 

\subsection{Covariate effects}\label{sec:Effects}

Here we report the covariate effects from the Bayesian GAM framework we followed after selecting our parameter space via LASSO. More specifically, out of the six covariates mentioned above, we opted to model the \emph{Slope$_\sigma$} and \emph{Soil Depth} as ordinal properties with a Random Walk of the first order used to impose adjacent class dependency. 

\begin{figure}[t!] 
	\centering
	\includegraphics[width=\linewidth]{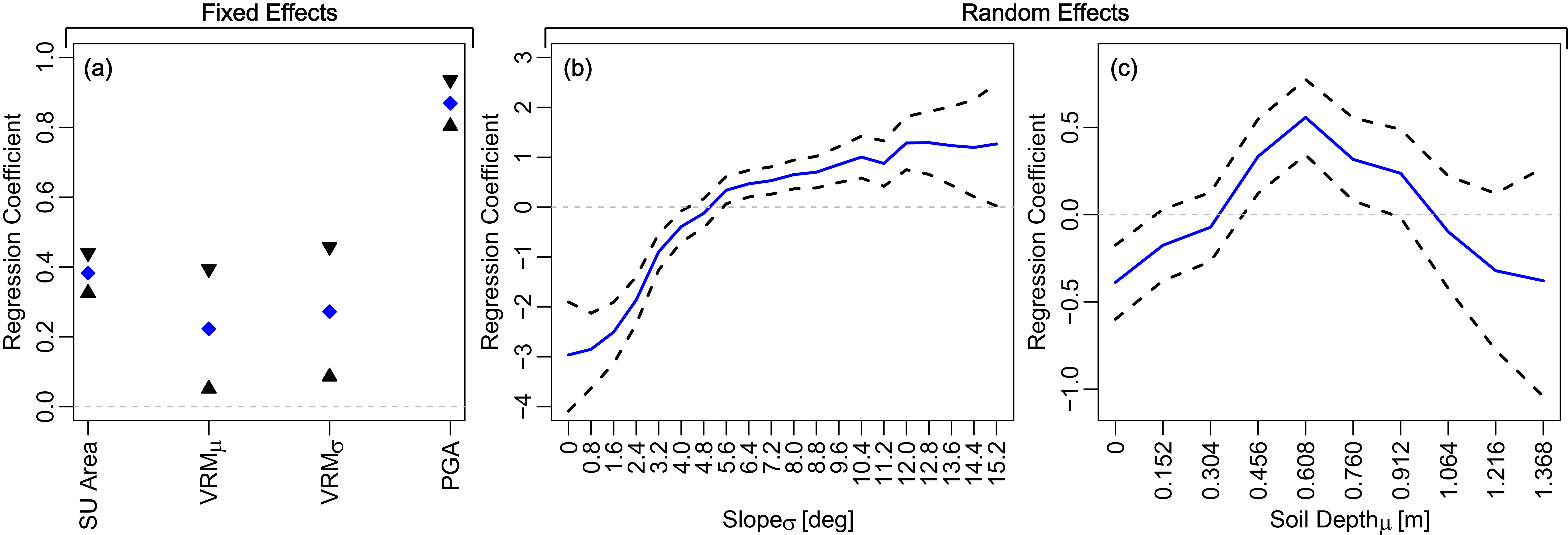}
	\caption{Summary of fixed and random effects of our reference susceptibility model for the Northridge earthquake. Fixed effects are shown in panel (a), whereas the random effect for the \emph{Slope$_\sigma$} is shown in panel (b) and the same for \emph{Soil Depth} is shown in panel (c).}
	\label{Figure4}
\end{figure}

Figure \ref{Figure4}a summarizes the posterior distribution of each covariate used linearly in our model. Specifically, they are all shown to be significant (each distribution is above the zero line) but with various levels of associated uncertainty. The \emph{SU Area} is clearly the parameter with the narrowest posterior distribution, followed by the \emph{PGA}. The two components of the \emph{VRM} signal have comparable uncertainty levels. Taking this aside, the \emph{PGA} shows the greatest influence with respect to the final susceptibility, for the regression coefficients are all expressed in the same unitless scale due to the mean zero unit variance step we mentioned in Section \ref{sec:Covars}. As for the two random effects, both clearly behave in a nonlinear fashion, supporting our choice. For the \emph{Slope$_\sigma$}, small variations of slope steepness within a SU negatively contribute to the susceptibility. From slope steepness variations (measured in $1\sigma$) greater than 5 degrees, the effect becomes positive, indicating landslide-prone conditions. As for the \emph{Soil Depth}, a peculiar pattern is shown. Two separate portions of the soil depth distribution contribute to decrease the estimated susceptibility whereas from approximately $0.4m$ to $1.0m$ the effect is opposite. This result well aligns to the description of the landslides provided by \citet{harp1995inventory,harp1996landslides}. The authors report that the vast majority of co-seismic landslides were shallow-seated.     

\subsection{Cross-validation performance}\label{sec:CV}

\begin{figure}[t!] 
	\centering
	\includegraphics[width=0.5\linewidth]{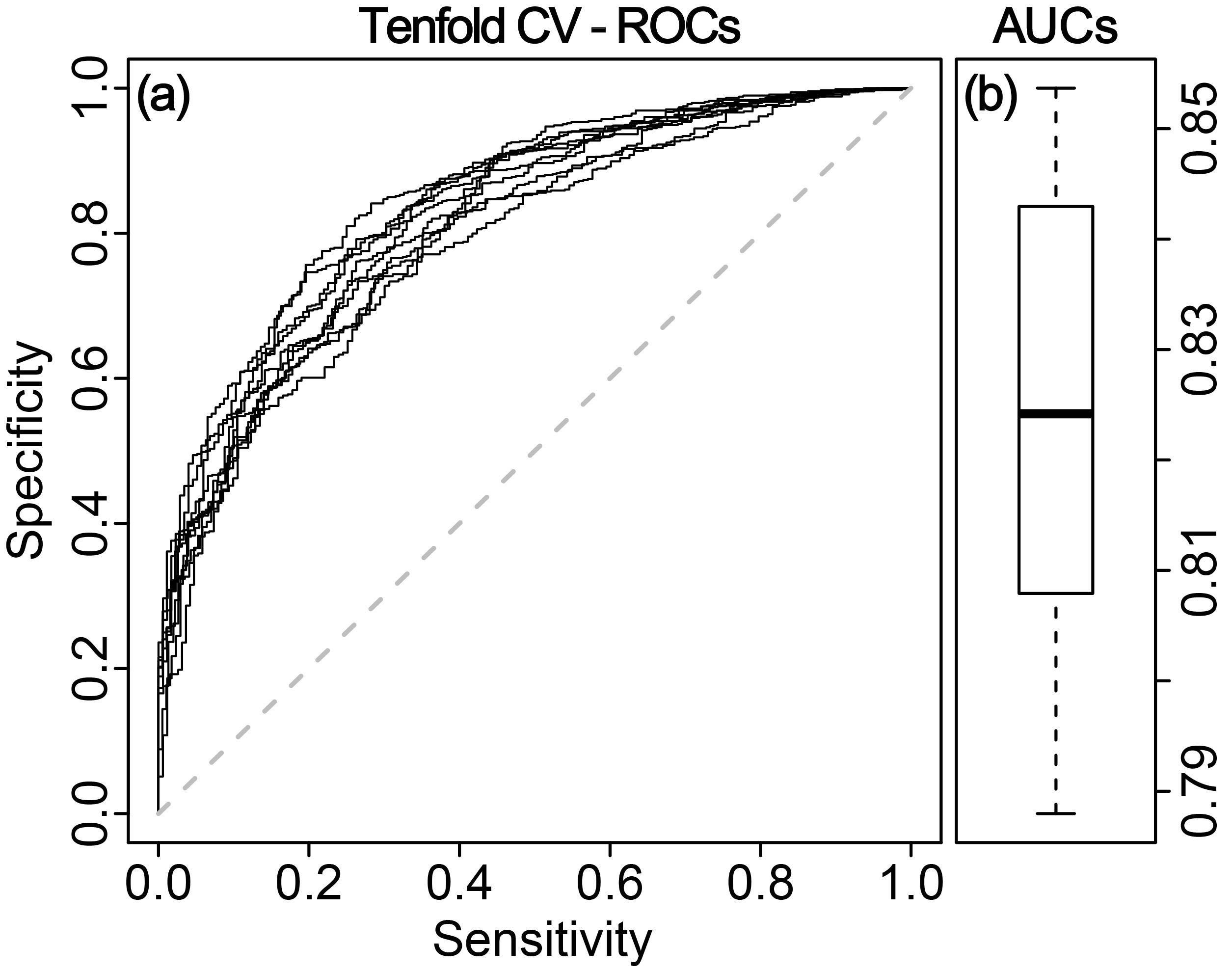}
	\caption{Summary the 10-fold CV performance. Panel (a) shows the Receiver Operating Characteristic curves, one for each of the ten CV. Panel (b) summarises the distribution of associated AUC values.}
	\label{Figure5}
\end{figure}

Figure \ref{Figure5} summarizes the overall predictive performance for the 10-fold CV scheme we implement for the co-Northridge susceptibility. As the validation subset changes, a limited variability in the resulted performance metrics can be seen. This is graphically depicted with a limited spread of the ROC curves and consequently in their AUC estimates. More specifically, The median AUC is approximately 0.82, which corresponds to an excellent performance according to the AUC classification proposed by \citet{hosmer2000}. 

\subsection{Benchmark Northridge susceptibility}\label{sec:NorthridgeSusc}

A number of studies have already focused on the co-seismic landslide susceptibility induced by the 1994 Northridge earthquake \citep[e.g.,][]{godt2008rapid,Nowicki2014,nowicki2018global,tanyas2019}. For this reason, and in light to the different aim of this work, we choose to report the posterior mean susceptibility in its numerical form, together with its uncertainty measured in a 95\% CI. For conciseness, we do not report the same information in map form. 

\begin{figure}[t!] 
	\centering
	\includegraphics[width=0.5\linewidth]{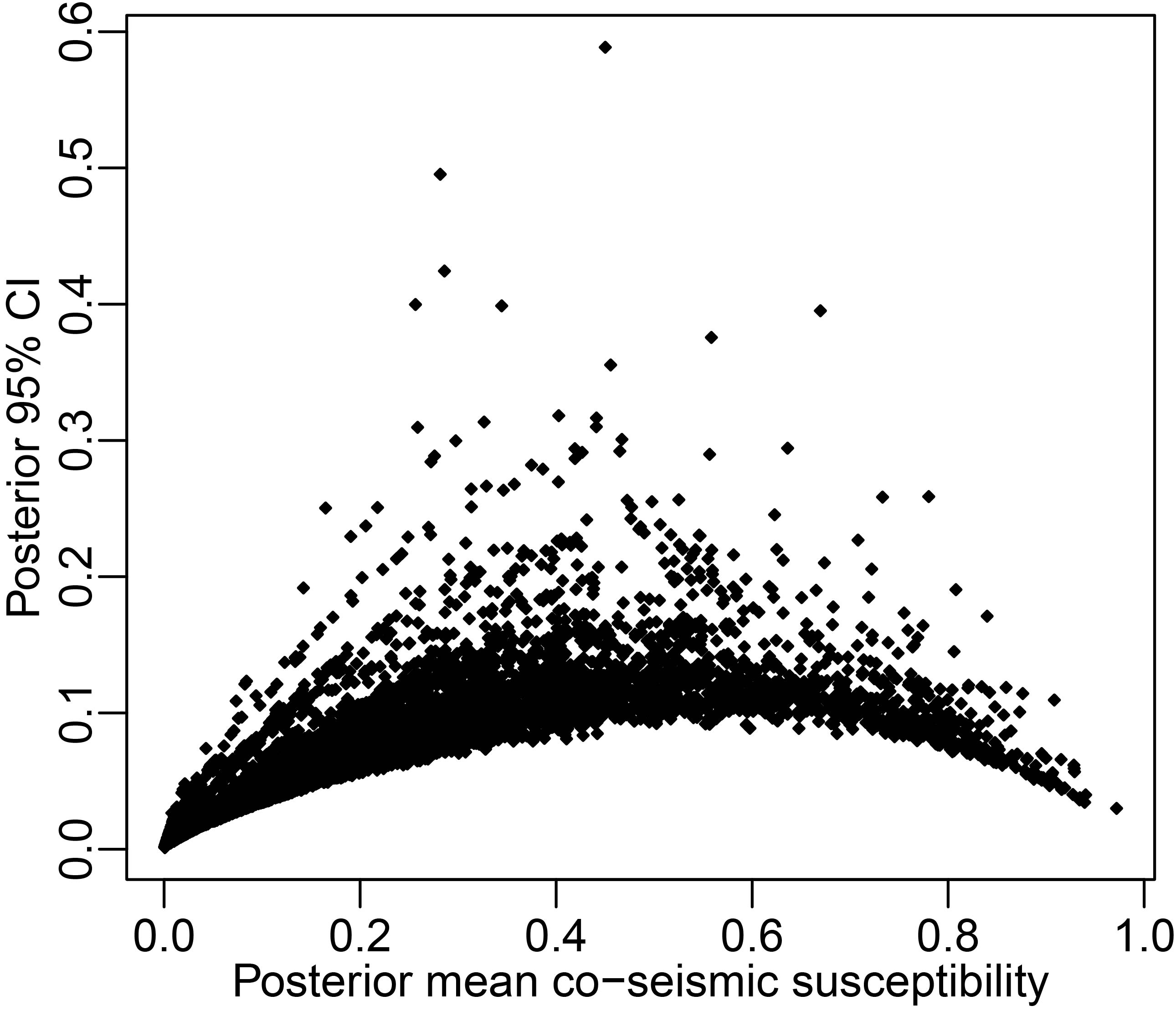}
	\caption{Error plot showing the relation between the posterior mean of the co-Northridge susceptibility and its associated uncertainty.}
	\label{Figure6}
\end{figure}

Figure \ref{Figure6} shows the error plot. This graph is crucial in most landslide susceptibility studies \citep{rossi2010}. In fact, an ideal model should produce very low susceptibility and very high susceptibility estimates associated with low uncertainty. In this case, the potential users of the susceptibility model can trust the result and plan accordingly. The only portion of the error plot which is accepted to report the highest uncertainty corresponds to the central portion of the susceptibility distribution, where the determination of likely landslide-prone SUs is intrinsically more difficult \citep{lombardo2016a,reichenbach2018}.  In this sense, the posterior estimates show a clear bell shape.

\subsection{Scenario-based susceptibility}\label{sec:ScenarioSusc}

\begin{figure}[t!] 
	\centering
	\includegraphics[width=0.6\linewidth]{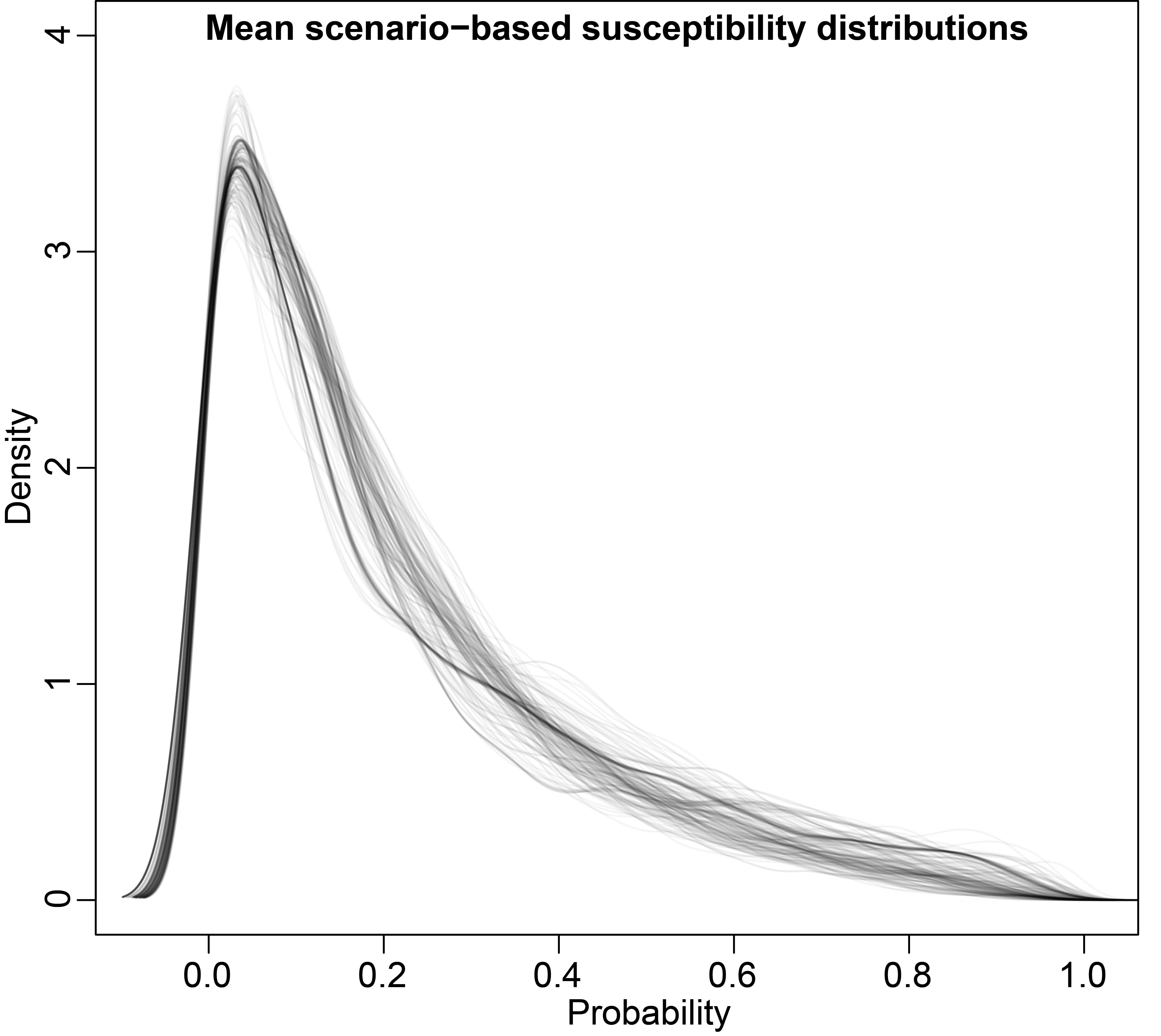}
	\caption{PDFs of the mean susceptibility estimated computed from the 1000 simulation batch for each of the 217 ground motion scenarios.}
	\label{Figure7}
\end{figure}

In this section we report the result of our simulation scheme. We remind here that we have run 1000 simulations for each of the 217 ground-motions scenarios extracted for the study area. This information is shared in the supplementary material (both SU shapefile and scenario-based susceptibility maps reported as the mean and 95\%CI for each of the 1000 simulation batches). In Figure \ref{Figure7} we show the Probability Density Functions of each mean scenario-based susceptibility computed from the ($1000\times217$) simulations.  

\begin{figure}[t!] 
	\centering
	\includegraphics[width=0.5\linewidth]{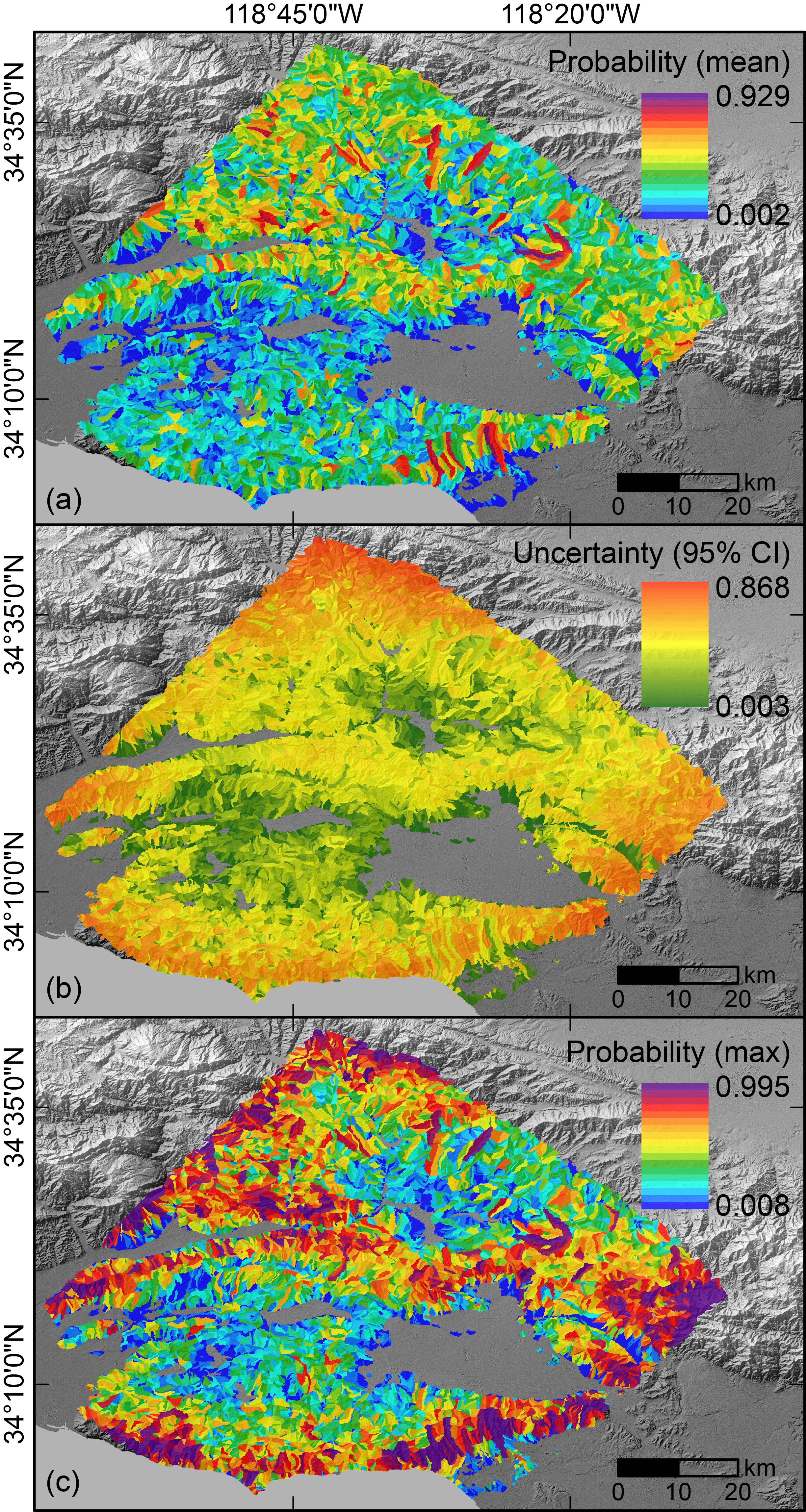}
	\caption{Descriptive statistics of the 217 ground motion scenarios (and 1000 simulations for each case) together with the 1994 Northridge susceptibility. Panel (a) shows the mean susceptibility of all the possible ground motions scenarios combined. Panel (b) reports the uncertainty in the susceptibility estimates for all scenarios. Panel (c) depicts the susceptibility corresponding to the maximum probability values for each SU and for all scenarios together.}
	\label{Figure8}
\end{figure}

A similar information is summarized in map form in Figure \ref{Figure8}, where out of all the simulations we extract the global mean and associated uncertainty. Additionally we report the worst-case scenario where we computed for each SU in the study area the maximum probability out of all the mean scenario-based susceptibilities. 

\begin{figure}[t!] 
	\centering
	\includegraphics[width=\linewidth]{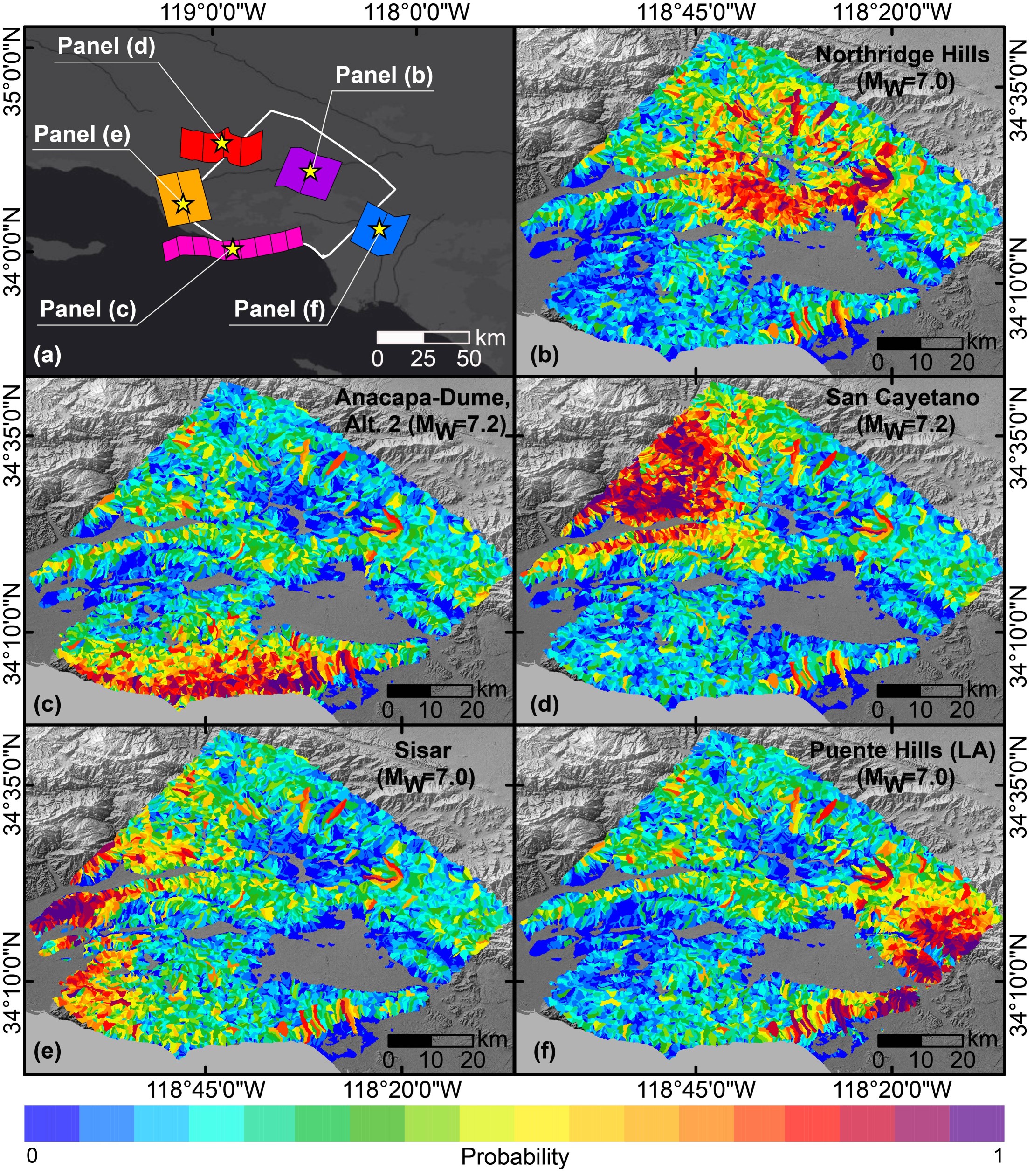}
	\caption{Panel (a) shows a geographical summary of a selected number of scenario bases epicenters and associated rupture planes. The white polygon corresponds to our study area. For each case shown in the previous panel, we report the corresponding mean susceptibility map out of the 1000 simulation runs (panels b to f).}
	\label{Figure9}
\end{figure}

To provide a more intuitive understanding of our simulation scheme we propose, we select five among the largest scenario-based ground motions available at the USGS Shakemap service, falling within the study area. Their epicenters, rupture planes and metadata are shown in Figure \ref{Figure9}a. In the same figure we plot the mean susceptibility map computed from the 1000 simulations, for each of the five scenarios (Figure \ref{Figure9}b-f).

\section{Discussions}\label{sec:Discuss}

In this work, we take advantage of a plug-in simulation procedure. We initially calibrate a GAM-susceptibility model for the 1994 Northridge earthquake. And, on the basis of the estimated fixed and random effects (and associated distributions) we substitute the PGA pattern from the Northridge with each of the available 217 scenarios within the study area. For each scenario, we then run 1000 simulations to capture the variability of the susceptibility pattern as the ground motion changes from one predicted scenario to another. 
The distribution of the mean susceptibility for each simulation batch appears quite different from case to case. This is shown in Figure \ref{Figure7} with large oscillations. This information is further clarified in Figure \ref{Figure8} where the mean susceptibility of all the $1000 \times 217$ simulations is plotted in map form together with the 95\% CI and maximum level. The global mean (Figure \ref{Figure8}a) shows few extremely susceptible slopes scattered over the landscape. Notably, by computing the mean for all simulations, the susceptibility may be smoothed and end up showing a low variability or a limited number of expected unstable SUs. Conversely, by looking at the maximum susceptibility across all the simulations (worst-case global scenario in Figure \ref{Figure8}c), three regions within the study area appear more prone to fail than the rest. The steep SUs located along the coastline are part of the worst case scenario as well as a sequence of SU approximately aligned from East to West. These are SUs exposed to the main bulk of scenario-based ground motion patterns. Ultimately, the third cluster of highly susceptible SUs is located to the Northern sector of the study area where the topography is significantly rough and scenario-based ground motion quite common among the selected cases.

To provide further examples of our simulation procedure, Figure \ref{Figure9} moves away from the global summary provided above, and depicts five cases for which the PGA was reported to the be among the largest of the 217 scenarios we examined. Here the ground motion effect is particularly evident for each of the five mean simulated susceptibilities reflects different susceptible SUs. However, despite we managed to capture the seismic effect over the co-Nortridge landslides and project it over the scenario-based ShakeMaps, the model may still lacks part of the information required to fully describe the physics of earthquake-induced ground deformation. In fact, on the one hand the directivity is not included in the ground motion scenarios \citep{Allstadt2018}. Therefore, some specific portion of the landscape aligned to the main direction of the earthquake radiation pattern may still not appear as susceptible as they may be. On the other hand, even the topographic amplification is not accounted for in the earthquake scenarios \citep{Allstadt2018}. As a result, the spatial dependence that narrow mountaintops should exhibit between susceptibility and the wavefield amplification is also neglected.    

However, despite minor spatial dependency issues, our model is able to efficiently include any PGA effect into any potential susceptibility realization within the study area under the assumption that the PGA effect we estimated for the Northridge model (see Section \ref{sec:NorthridgeSusc}) is robust. We believe so because the governing mechanical laws that link the slope instability to the ground shaking should not change with time. As a result, the only limiting factor for similar procedures should depend on the statistical inference. In this sense, we believe the fixed effect for the PGA to be well-estimated (see Figure \ref{Figure4}a) both in its amplitude (posterior mean) and significance (posterior 95\% credible interval above the zero line).     

\section{Concluding remarks}\label{sec:Conclusions}

Throughout the manuscript we have referred to our simulation results in terms of susceptibility. However, we would like to point out that the scenario-based PGA estimates provided by the USGS ShakeMap system have a temporal connotation. As mentioned before, \citep{USGS2017} report each scenario to be associated with a return time of 50 years. Therefore, our results, either if we examine specific scenarios like in Figure \ref{Figure9} or if we consider the combined effects of all scenarios together like in Figure \ref{Figure8}, both express probabilities of landslide occurrence (susceptibility) within an expected timespan of 50 years. This definition satisfies most the notion of landslide hazard \cite{fell2008guidelines, guzzetti1999landslidehazard}. As a result, we argue that the method we propose allows one to developed scenario-based landslide-hazard maps.  

We also stress here that the use of the scenario-based ground motion is only one possible application in simulating landslide landslide prone conditions. In fact, one could follow the same strategy and simulate over precipitations, land use changes and or, much closer to the framework we propose here, over past ground motion scenarios. The latter could unravel potential landslide prone-conditions in the past and help explore historical data for which data is usually poor or non-existing. An example that goes in this research direction can be found in the companion paper to the present one \citep{Luo2020}.      

\bibliographystyle{CUP}
\bibliography{landslides}

\begin{thebibliography}{89}

\bibitem[Abrahamson and Bommer(2005)]{abrahamson2005}
Abrahamson, N.~A. and Bommer, J.~J. (2005) Probability and uncertainty in
  seismic hazard analysis.
\newblock \emph{Earthquake spectra} \textbf{21}(2), 603--607.

\bibitem[Akg{\"u}n and T{\"u}rk(2011)]{akgun2011}
Akg{\"u}n, A. and T{\"u}rk, N. (2011) {Mapping erosion susceptibility by a
  multivariate statistical method: a case study from the Ayval{\i}k region, NW
  Turkey}.
\newblock \emph{Computers \& geosciences} \textbf{37}(9), 1515--1524.

\bibitem[Aleotti(2004)]{aleotti2004}
Aleotti, P. (2004) A warning system for rainfall-induced shallow failures.
\newblock \emph{Engineering geology} \textbf{73}(3-4), 247--265.

\bibitem[Allstadt \emph{et~al.}(2018)Allstadt, Jibson, Thompson, Massey, Wald,
  Godt and Rengers]{Allstadt2018}
Allstadt, K.~E., Jibson, R.~W., Thompson, E.~M., Massey, C.~I., Wald, D.~J.,
  Godt, J.~W. and Rengers, F.~K. (2018) Improving {N}ear-{R}eal-{T}ime
  {C}oseismic {L}andslide {M}odels: {L}essons {L}earned from the 2016
  {K}aik{\=o}ura, {N}ew {Z}ealand, {E}arthquake.
\newblock \emph{Bulletin of the Seismological Society of America} .

\bibitem[Allstadt \emph{et~al.}(2017)Allstadt, Thompson, Hearne, Jessee, Zhu,
  Wald and Tanyas]{allstadt2017integrating}
Allstadt, K.~E., Thompson, E.~M., Hearne, M., Jessee, M.~N., Zhu, J., Wald,
  D.~J. and Tanyas, H. (2017) {Integrating landslide and liquefaction hazard
  and loss estimates with existing USGS real-time earthquake information
  products}.
\newblock In \emph{16th World Conference on Earthquake Engineering}.

\bibitem[Alvioli \emph{et~al.}(2016)Alvioli, Marchesini, Reichenbach, Rossi,
  Ardizzone, Fiorucci and Guzzetti]{alvioli2016automatic}
Alvioli, M., Marchesini, I., Reichenbach, P., Rossi, M., Ardizzone, F.,
  Fiorucci, F. and Guzzetti, F. (2016) Automatic delineation of
  geomorphological slope units with r.slopeunits v1.0 and their optimization
  for landslide susceptibility modeling.
\newblock \emph{Geoscientific Model Development} \textbf{9}(11), 3975--3991.

\bibitem[Amato \emph{et~al.}(2019)Amato, Eisank, Castro-Camilo and
  Lombardo]{amato2019accounting}
Amato, G., Eisank, C., Castro-Camilo, D. and Lombardo, L. (2019) Accounting for
  covariate distributions in slope-unit-based landslide susceptibility models.
  a case study in the alpine environment.
\newblock \emph{Engineering Geology} \textbf{260}, In print.

\bibitem[Arabameri \emph{et~al.}(2020)Arabameri, Saha, Roy, Chen, Blaschke and
  Tien~Bui]{arabameri2020}
Arabameri, A., Saha, S., Roy, J., Chen, W., Blaschke, T. and Tien~Bui, D.
  (2020) {Landslide Susceptibility Evaluation and Management Using Different
  Machine Learning Methods in The Gallicash River Watershed, Iran}.
\newblock \emph{Remote Sensing} \textbf{12}(3), 475.

\bibitem[Bakka \emph{et~al.}(2018)Bakka, Rue, Fuglstad, Riebler, Bolin,
  Krainski, Simpson and Lindgren]{Bakka.etal:2018}
Bakka, H., Rue, H., Fuglstad, G.~A., Riebler, A., Bolin, D., Krainski, E.,
  Simpson, D. and Lindgren, F. (2018) Spatial modelling with r-inla: A review.
\newblock Submitted, arXiv preprint:1802.06350.

\bibitem[Brenning(2005)]{brenning2005spatial}
Brenning, A. (2005) Spatial prediction models for landslide hazards: review,
  comparison and evaluation.
\newblock \emph{Natural Hazards and Earth System Science} \textbf{5}(6),
  853--862.

\bibitem[Brenning(2008)]{brenning2008}
Brenning, A. (2008) {Statistical geocomputing combining R and SAGA: The example
  of landslide susceptibility analysis with generalized additive models}.
\newblock \emph{Hamburger Beitr{\"a}ge zur Physischen Geographie und
  Landschafts{\"o}kologie} \textbf{19}(23-32), 410.

\bibitem[Budimir \emph{et~al.}(2015)Budimir, Atkinson and
  Lewis]{budimir-SystematicReview-2015}
Budimir, M. E.~A., Atkinson, P.~M. and Lewis, H.~G. (2015) A systematic review
  of landslide probability mapping using logistic regression.
\newblock \emph{Landslides} pp. 1--18.

\bibitem[Cama \emph{et~al.}(2015)Cama, Lombardo, Conoscenti, Agnesi and
  Rotigliano]{cama2015predicting}
Cama, M., Lombardo, L., Conoscenti, C., Agnesi, V. and Rotigliano, E. (2015)
  Predicting storm-triggered debris flow events: application to the 2009
  {I}onian {P}eloritan disaster ({S}icily, {I}taly).
\newblock \emph{Nat Hazards Earth Syst Sci} \textbf{15}(8), 1785--1806.

\bibitem[Cama \emph{et~al.}(2017)Cama, Lombardo, Conoscenti and
  Rotigliano]{cama2017improving}
Cama, M., Lombardo, L., Conoscenti, C. and Rotigliano, E. (2017) Improving
  transferability strategies for debris flow susceptibility assessment:
  {A}pplication to the {S}aponara and {I}tala catchments ({M}essina, {I}taly).
\newblock \emph{Geomorphology} \textbf{288}, 52--65.

\bibitem[Castro~Camilo \emph{et~al.}(2017)Castro~Camilo, Lombardo, Mai, Dou and
  Huser]{Camilo2017}
Castro~Camilo, D., Lombardo, L., Mai, P., Dou, J. and Huser, R. (2017) Handling
  high predictor dimensionality in slope-unit-based landslide susceptibility
  models through {LASSO}-penalized {G}eneralized {L}inear {M}odel.
\newblock \emph{Environmental Modelling and Software} \textbf{97}, 145--156.

\bibitem[Conoscenti \emph{et~al.}(2016)Conoscenti, Rotigliano, Cama,
  Caraballo-Arias, Lombardo and Agnesi]{conoscenti2016exploring}
Conoscenti, C., Rotigliano, E., Cama, M., Caraballo-Arias, N.~A., Lombardo, L.
  and Agnesi, V. (2016) Exploring the effect of absence selection on landslide
  susceptibility models: A case study in {S}icily, {I}taly.
\newblock \emph{Geomorphology} \textbf{261}, 222--235.

\bibitem[Das \emph{et~al.}(2010)Das, Sahoo, van Westen, Stein and
  Hack]{das2010}
Das, I., Sahoo, S., van Westen, C., Stein, A. and Hack, R. (2010) {Landslide
  susceptibility assessment using logistic regression and its comparison with a
  rock mass classification system, along a road section in the northern
  Himalayas (India)}.
\newblock \emph{Geomorphology} \textbf{114}(4), 627--637.

\bibitem[Del~Gaudio \emph{et~al.}(2003)Del~Gaudio, Pierri and
  Wasowski]{gaudio2003approach}
Del~Gaudio, V., Pierri, P. and Wasowski, J. (2003) {An approach to
  time-probabilistic evaluation of seismically induced landslide hazard}.
\newblock \emph{Bulletin of the Seismological Society of America}
  \textbf{93}(2), 557--569.

\bibitem[Del~Gaudio and Wasowski(2004)]{gaudio2004time}
Del~Gaudio, V. and Wasowski, J. (2004) {Time probabilistic evaluation of
  seismically induced landslide hazard in Irpinia (Southern Italy)}.
\newblock \emph{Soil Dynamics and Earthquake Engineering} \textbf{24}(12),
  915--928.

\bibitem[Fan \emph{et~al.}(2019)Fan, Scaringi, Korup, West, van Westen, Tanyas,
  Hovius, Hales, Jibson, Allstadt \emph{et~al.}]{fan2019}
Fan, X., Scaringi, G., Korup, O., West, A.~J., van Westen, C.~J., Tanyas, H.,
  Hovius, N., Hales, T.~C., Jibson, R.~W., Allstadt, K.~E. \emph{et~al.} (2019)
  {Earthquake-induced chains of geologic hazards: Patterns, mechanisms, and
  impacts}.
\newblock \emph{Reviews of geophysics} \textbf{57}(2), 421--503.

\bibitem[Fell \emph{et~al.}(2008)Fell, Corominas, Bonnard, Cascini, Leroi,
  Savage \emph{et~al.}]{fell2008guidelines}
Fell, R., Corominas, J., Bonnard, C., Cascini, L., Leroi, E., Savage, W.~Z.
  \emph{et~al.} (2008) Guidelines for landslide susceptibility, hazard and risk
  zoning for land-use planning.
\newblock \emph{Engineering Geology} \textbf{102}(3-4), 99--111.

\bibitem[Ferkingstad \emph{et~al.}(2015)Ferkingstad, Rue
  \emph{et~al.}]{ferkingstad2015}
Ferkingstad, E., Rue, H. \emph{et~al.} (2015) {Improving the INLA approach for
  approximate Bayesian inference for latent Gaussian models}.
\newblock \emph{Electronic Journal of Statistics} \textbf{9}(2), 2706--2731.

\bibitem[Friedman \emph{et~al.}(2009)Friedman, Hastie and
  Tibshirani]{friedman2009}
Friedman, J., Hastie, T. and Tibshirani, R. (2009) {glmnet: Lasso and
  elastic-net regularized generalized linear models}.
\newblock \emph{R package version} \textbf{1}(4).

\bibitem[Ghosh \emph{et~al.}(2012)Ghosh, van Westen, Carranza, Jetten,
  Cardinali, Rossi and Guzzetti]{ghosh2012}
Ghosh, S., van Westen, C.~J., Carranza, E. J.~M., Jetten, V.~G., Cardinali, M.,
  Rossi, M. and Guzzetti, F. (2012) {Generating event-based landslide maps in a
  data-scarce Himalayan environment for estimating temporal and magnitude
  probabilities}.
\newblock \emph{Engineering geology} \textbf{128}, 49--62.

\bibitem[Godt \emph{et~al.}(2008)Godt, Sener, Verdin, Wald, Earle, Harp and
  Jibson]{godt2008rapid}
Godt, J., Sener, B., Verdin, K., Wald, D., Earle, P., Harp, E. and Jibson, R.
  (2008) {Rapid assessment of earthquake-induced landsliding}.
\newblock In \emph{Proceedings of the First World Landslide Forum, United
  Nations University, Tokyo}, volume~4, pp. 3166--1.

\bibitem[Goetz \emph{et~al.}(2015)Goetz, Brenning, Petschko and
  Leopold]{goetz2015}
Goetz, J., Brenning, A., Petschko, H. and Leopold, P. (2015) Evaluating machine
  learning and statistical prediction techniques for landslide susceptibility
  modeling.
\newblock \emph{Computers \& geosciences} \textbf{81}, 1--11.

\bibitem[Graziella \emph{et~al.}(2015)Graziella, Ingeborg, Monica,
  Nils-Kristian, Ragnar, Erik and Herv{\'e}]{devoli2015}
Graziella, D., Ingeborg, K., Monica, S., Nils-Kristian, O., Ragnar, E., Erik,
  J. and Herv{\'e}, C. (2015) {Landslide early warning system and web tools for
  real-time scenarios and for distribution of warning messages in Norway}.
\newblock In \emph{Engineering Geology for Society and Territory-Volume 2}, pp.
  625--629. Springer.

\bibitem[Greco \emph{et~al.}(2013)Greco, Giorgio, Capparelli and
  Versace]{greco2013}
Greco, R., Giorgio, M., Capparelli, G. and Versace, P. (2013) Early warning of
  rainfall-induced landslides based on empirical mobility function predictor.
\newblock \emph{Engineering Geology} \textbf{153}, 68--79.

\bibitem[Guzzetti \emph{et~al.}(2004)Guzzetti, Cardinali, Reichenbach, Cipolla,
  Sebastiani, Galli and Salvati]{guzzetti2004}
Guzzetti, F., Cardinali, M., Reichenbach, P., Cipolla, F., Sebastiani, C.,
  Galli, M. and Salvati, P. (2004) {Landslides triggered by the 23 November
  2000 rainfall event in the Imperia Province, Western Liguria, Italy}.
\newblock \emph{Engineering Geology} \textbf{73}(3-4), 229--245.

\bibitem[Guzzetti \emph{et~al.}(1999)Guzzetti, Carrara, Cardinali and
  Reichenbach]{guzzetti1999landslidehazard}
Guzzetti, F., Carrara, A., Cardinali, M. and Reichenbach, P. (1999) {Landslide
  Hazard Evaluation: A Review of Current Techniques and Their Application in a
  Multi-Scale Study, Central Italy}.
\newblock \emph{Geomorphology} \textbf{31}(1), 181--216.

\bibitem[Guzzetti \emph{et~al.}(2019)Guzzetti, Gariano, Peruccacci, Brunetti,
  Marchesini, Rossi and Melillo]{guzzetti2019}
Guzzetti, F., Gariano, S.~L., Peruccacci, S., Brunetti, M.~T., Marchesini, I.,
  Rossi, M. and Melillo, M. (2019) Geographical landslide early warning
  systems.
\newblock \emph{Earth-Science Reviews} p. 102973.

\bibitem[Hanks \emph{et~al.}(2005)Hanks, Abrahamson, Board, Boore, Brune and
  Cornell]{hanks2005}
Hanks, T.~C., Abrahamson, N., Board, M., Boore, D.~M., Brune, J. and Cornell,
  C. (2005) Observed ground motions, extreme ground motions, and physical
  limits to ground motions.
\newblock In \emph{Directions in Strong Motion Instrumentation}, pp. 55--59.
  Springer.

\bibitem[Harp and Jibson(1995)]{harp1995inventory}
Harp, E.~L. and Jibson, R.~W. (1995) {Inventory of landslides triggered by the
  1994 Northridge, California earthquake}.
\newblock Technical report, US Geological Survey,.

\bibitem[Harp and Jibson(1996)]{harp1996landslides}
Harp, E.~L. and Jibson, R.~W. (1996) {Landslides triggered by the 1994
  Northridge, California, earthquake}.
\newblock \emph{Bulletin of the Seismological Society of America}
  \textbf{86}(1B), S319--S332.

\bibitem[Harp \emph{et~al.}(2011)Harp, Keefer, Sato and
  Yagi]{harp2011landslide}
Harp, E.~L., Keefer, D.~K., Sato, H.~P. and Yagi, H. (2011) {Landslide
  inventories: the essential part of seismic landslide hazard analyses}.
\newblock \emph{Engineering Geology} \textbf{122}(1-2), 9--21.

\bibitem[Harrell~Jr(2015)]{harrell2015}
Harrell~Jr, F.~E. (2015) \emph{Regression modeling strategies: with
  applications to linear models, logistic and ordinal regression, and survival
  analysis}.
\newblock Springer.

\bibitem[Hastie and Qian(2014)]{hastie2014}
Hastie, T. and Qian, J. (2014) Glmnet vignette.
\newblock \emph{Retrieve from http://www. web. stanford. edu/\~{}
  hastie/Papers/Glmnet\_Vignette. pdf. Accessed September} \textbf{20}, 2016.

\bibitem[Heerdegen and Beran(1982)]{heerdegen1982quantifying}
Heerdegen, R.~G. and Beran, M.~A. (1982) Quantifying source areas through land
  surface curvature and shape.
\newblock \emph{Journal of Hydrology} \textbf{57}(3-4), 359--373.

\bibitem[Hosmer and Lemeshow(2000)]{hosmer2000}
Hosmer, D.~W. and Lemeshow, S. (2000) \emph{{Applied Logistic Regression}}.
  Second edition.
\newblock New York: Wiley.

\bibitem[Jarvis \emph{et~al.}(2008)Jarvis, Reuter, Nelson and
  Guevara]{jarvis2008hole}
Jarvis, A., Reuter, H.~I., Nelson, A. and Guevara, E. (2008) Hole-filled {SRTM}
  for the globe {V}ersion 4 .

\bibitem[Jasiewicz and Stepinski(2013)]{jasiewicz2013geomorphons}
Jasiewicz, J. and Stepinski, T.~F. (2013) {Geomorphons—a pattern recognition
  approach to classification and mapping of landforms}.
\newblock \emph{Geomorphology} \textbf{182}, 147--156.

\bibitem[Jibson \emph{et~al.}(1998)Jibson, Harp and Michael]{jibson1998method}
Jibson, R.~W., Harp, E.~L. and Michael, J.~A. (1998) \emph{{A method for
  producing digital probabilistic seismic landslide hazard maps: an example
  from the Los Angeles, California, area}}.
\newblock US Department of the Interior, US Geological Survey.

\bibitem[Jibson \emph{et~al.}(2000)Jibson, Harp and Michael]{jibson2000method}
Jibson, R.~W., Harp, E.~L. and Michael, J.~A. (2000) {A method for producing
  digital probabilistic seismic landslide hazard maps}.
\newblock \emph{Engineering Geology} \textbf{58}(3-4), 271--289.

\bibitem[Kirschbaum \emph{et~al.}(2009)Kirschbaum, Adler, Hong and
  Lerner-Lam]{kirschbaum2009}
Kirschbaum, D., Adler, R., Hong, Y. and Lerner-Lam, A. (2009) Evaluation of a
  preliminary satellite-based landslide hazard algorithm using global landslide
  inventories.
\newblock \emph{Natural Hazards and Earth System Sciences} \textbf{9}(3),
  673--686.

\bibitem[Kirschbaum and Stanley(2018)]{kirschbaum2018satellite}
Kirschbaum, D. and Stanley, T. (2018) {Satellite-Based Assessment of
  Rainfall-Triggered Landslide Hazard for Situational Awareness}.
\newblock \emph{Earth's Future} \textbf{6}(3), 505--523.

\bibitem[Kirschbaum \emph{et~al.}(2015)Kirschbaum, Stanley and
  Zhou]{kirschbaum2015}
Kirschbaum, D., Stanley, T. and Zhou, Y. (2015) Spatial and temporal analysis
  of a global landslide catalog.
\newblock \emph{Geomorphology} \textbf{249}, 4--15.

\bibitem[Kirschbaum \emph{et~al.}(2010)Kirschbaum, Adler, Hong, Hill and
  Lerner-Lam]{kirschbaum2010}
Kirschbaum, D.~B., Adler, R., Hong, Y., Hill, S. and Lerner-Lam, A. (2010) A
  global landslide catalog for hazard applications: method, results, and
  limitations.
\newblock \emph{Natural Hazards} \textbf{52}(3), 561--575.

\bibitem[Ko and Lo(2018)]{ko2018}
Ko, F.~W. and Lo, F.~L. (2018) {From landslide susceptibility to landslide
  frequency: A territory--wide study in Hong Kong}.
\newblock \emph{Engineering geology} \textbf{242}, 12--22.

\bibitem[Lari \emph{et~al.}(2014)Lari, Frattini and Crosta]{lari2014}
Lari, S., Frattini, P. and Crosta, G. (2014) A probabilistic approach for
  landslide hazard analysis.
\newblock \emph{Engineering geology} \textbf{182}, 3--14.

\bibitem[Lee \emph{et~al.}(2000)Lee, Lee and Han]{lee2000}
Lee, L.~H., Lee, H.~H. and Han, S.~W. (2000) Method of selecting design
  earthquake ground motions for tall buildings.
\newblock \emph{The structural design of tall buildings} \textbf{9}(3),
  201--213.

\bibitem[Lombardo \emph{et~al.}(2016)Lombardo, Fubelli, Amato and
  Bonasera]{lombardo2016a}
Lombardo, L., Fubelli, G., Amato, G. and Bonasera, M. (2016) Presence-only
  approach to assess landslide triggering-thickness susceptibility: a test for
  the {M}ili catchment (north-eastern {S}icily, {I}taly).
\newblock \emph{Natural Hazards} \textbf{84}(1), 565--588.

\bibitem[Lombardo and Mai(2018)]{lombardo2018presenting}
Lombardo, L. and Mai, P.~M. (2018) Presenting logistic regression-based
  landslide susceptibility results.
\newblock \emph{Engineering geology} \textbf{244}, 14--24.

\bibitem[Lombardo \emph{et~al.}(2019{a})Lombardo, Opitz, Ardizzone, Guzzetti
  and Huser]{lombardo2019space}
Lombardo, L., Opitz, T., Ardizzone, F., Guzzetti, F. and Huser, R. (2019{a})
  {Space-Time Landslide Predictive Modelling}.
\newblock \emph{arXiv preprint arXiv:1912.01233} .

\bibitem[Lombardo \emph{et~al.}(2018)Lombardo, Opitz and
  Huser]{Lombardo.etal:2018}
Lombardo, L., Opitz, T. and Huser, R. (2018) Point process-based modeling of
  multiple debris flow landslides using {INLA}: an application to the 2009
  {M}essina disaster.
\newblock \emph{Stochastic Environmental Research and Risk Assessment}
  \textbf{32}(7), 2179--2198.

\bibitem[Lombardo \emph{et~al.}(2019{b})Lombardo, Opitz and
  Huser]{lombardo2019numerical}
Lombardo, L., Opitz, T. and Huser, R. (2019{b}) {3 - Numerical Recipes for
  Landslide Spatial Prediction Using R-INLA: A Step-by-Step Tutorial}.
\newblock In \emph{Spatial Modeling in GIS and R for Earth and Environmental
  Sciences}, eds H.~R. Pourghasemi and C.~Gokceoglu, pp. 55 -- 83. Elsevier.
\newblock ISBN 978-0-12-815226-3.

\bibitem[Luo \emph{et~al.}(2020)Luo, Pei, Westen, Huang and Lombardo]{Luo2020}
Luo, L., Pei, X., Westen, C.~v., Huang, R. and Lombardo, L. (2020) {From
  scenario-based seismic hazard to scenario-based landslide hazard: rewinding
  to the past via statistical simulations}.
\newblock \emph{Submitted to Engineering Geology} .

\bibitem[Malamud \emph{et~al.}(2004)Malamud, Turcotte, Guzzetti and
  Reichenbach]{malamud2004}
Malamud, B.~D., Turcotte, D.~L., Guzzetti, F. and Reichenbach, P. (2004)
  Landslide inventories and their statistical properties.
\newblock \emph{Earth Surface Processes and Landforms} \textbf{29}(6),
  687--711.

\bibitem[Mathew \emph{et~al.}(2009)Mathew, Jha and Rawat]{mathew2009}
Mathew, J., Jha, V. and Rawat, G. (2009) {Landslide susceptibility zonation
  mapping and its validation in part of Garhwal Lesser Himalaya, India, using
  binary logistic regression analysis and receiver operating characteristic
  curve method}.
\newblock \emph{Landslides} \textbf{6}(1), 17--26.

\bibitem[Melchiorre and Frattini(2012)]{melchiorre2012}
Melchiorre, C. and Frattini, P. (2012) {Modelling probability of
  rainfall-induced shallow landslides in a changing climate, Otta, Central
  Norway}.
\newblock \emph{Climatic change} \textbf{113}(2), 413--436.

\bibitem[Montilla \emph{et~al.}(2003)Montilla, Hamdache and
  Casado]{montilla2003}
Montilla, J. A.~P., Hamdache, M. and Casado, C.~L. (2003) {Seismic hazard in
  Northern Algeria using spatially smoothed seismicity. Results for peak ground
  acceleration}.
\newblock \emph{Tectonophysics} \textbf{372}(1-2), 105--119.

\bibitem[Newmark(1965)]{newmark1965effects}
Newmark, N.~M. (1965) {Effects of earthquakes on dams and embankments}.
\newblock \emph{Geotechnique} \textbf{15}(2), 139--160.

\bibitem[Nowicki \emph{et~al.}(2014)Nowicki, Wald, Hamburger, Hearne and
  Thompson]{Nowicki2014}
Nowicki, M.~A., Wald, D.~J., Hamburger, M.~W., Hearne, M. and Thompson, E.~M.
  (2014) Development of a globally applicable model for near real-time
  prediction of seismically induced landslides.
\newblock \emph{Engineering Geology} \textbf{173}, 54--65.

\bibitem[Nowicki~Jessee \emph{et~al.}(2018)Nowicki~Jessee, Hamburger, Allstadt,
  Wald, Robeson, Tanyas, Hearne and Thompson]{nowicki2018global}
Nowicki~Jessee, M., Hamburger, M., Allstadt, K., Wald, D., Robeson, S., Tanyas,
  H., Hearne, M. and Thompson, E. (2018) {A Global Empirical Model for
  Near-Real-Time Assessment of Seismically Induced Landslides}.
\newblock \emph{Journal of Geophysical Research: Earth Surface}
  \textbf{123}(8), 1835--1859.

\bibitem[NRCS(2010)]{nrcs2010soil}
NRCS (2010) {Soil survey staff, natural resources conservation service, United
  States department of agriculture}.
\newblock \emph{Soil Survey Geographic (SSURGO) Database for northeast
  Tennessee.} .

\bibitem[Petersen \emph{et~al.}(2008)Petersen, Frankel, Harmsen, Mueller,
  Haller, Wheeler, Wesson, Zeng, Boyd, Perkins
  \emph{et~al.}]{petersen2008documentation}
Petersen, M.~D., Frankel, A.~D., Harmsen, S.~C., Mueller, C.~S., Haller, K.~M.,
  Wheeler, R.~L., Wesson, R.~L., Zeng, Y., Boyd, O.~S., Perkins, D.~M.
  \emph{et~al.} (2008) {Documentation for the 2008 update of the United States
  national seismic hazard maps}.
\newblock Technical report, Geological Survey (US).

\bibitem[Rathje and Saygili(2008)]{rathje2008probabilistic}
Rathje, E.~M. and Saygili, G. (2008) {Probabilistic seismic hazard analysis for
  the sliding displacement of slopes: scalar and vector approaches}.
\newblock \emph{Journal of Geotechnical and Geoenvironmental Engineering}
  \textbf{134}(6), 804--814.

\bibitem[Reichenbach \emph{et~al.}(2018)Reichenbach, Rossi, Malamud, Mihir and
  Guzzetti]{reichenbach2018}
Reichenbach, P., Rossi, M., Malamud, B.~D., Mihir, M. and Guzzetti, F. (2018) A
  review of statistically--based landslide susceptibility models.
\newblock \emph{Earth-Science Reviews} \textbf{180}, 60--91.

\bibitem[Romeo(2000)]{romeo2000seismically}
Romeo, R. (2000) {Seismically induced landslide displacements: a predictive
  model}.
\newblock \emph{Engineering Geology} \textbf{58}(3-4), 337--351.

\bibitem[Rossi \emph{et~al.}(2010)Rossi, Witt, Guzzetti, Malamud and
  Peruccacci]{rossi2010}
Rossi, M., Witt, A., Guzzetti, F., Malamud, B.~D. and Peruccacci, S. (2010)
  {Analysis of historical landslide time series in the Emilia--Romagna region,
  northern Italy}.
\newblock \emph{Earth Surface Processes and Landforms} \textbf{35}(10),
  1123--1137.

\bibitem[Rue \emph{et~al.}(2009)Rue, Martino and Chopin]{Rue2009}
Rue, H., Martino, S. and Chopin, N. (2009) Approximate {B}ayesian inference for
  latent {G}aussian models by using integrated nested {L}aplace approximations.
\newblock \emph{Journal of the Royal Statistical Society: Series B}
  \textbf{71}(2), 319--392.

\bibitem[Rue \emph{et~al.}(2017)Rue, Riebler, S{\o}rbye, Illian, Simpson and
  Lindgren]{Rue.al.2017}
Rue, H., Riebler, A., S{\o}rbye, S.~H., Illian, J.~B., Simpson, D.~P. and
  Lindgren, F.~K. (2017) Bayesian computing with {INLA}: A review.
\newblock \emph{Annual Review of Statistics and Its Application} \textbf{4},
  395--421.

\bibitem[Samia \emph{et~al.}(2017)Samia, Temme, Bregt, Wallinga, Guzzetti,
  Ardizzone and Rossi]{samia-LandslidesFollow-2017}
Samia, J., Temme, A.~J., Bregt, A., Wallinga, J., Guzzetti, F., Ardizzone, F.
  and Rossi, M. (2017) {Do Landslides Follow Landslides? Insights in Path
  Dependency from a Multi-Temporal Landslide Inventory}.
\newblock \emph{Landslides} \textbf{14}, 547--558.

\bibitem[Sappington \emph{et~al.}(2007)Sappington, Longshore and
  Thompson]{sappington2007quantifying}
Sappington, J.~M., Longshore, K.~M. and Thompson, D.~B. (2007) {Quantifying
  landscape ruggedness for animal habitat analysis: a case study using bighorn
  sheep in the Mojave Desert}.
\newblock \emph{The Journal of wildlife management} \textbf{71}(5), 1419--1426.

\bibitem[Saygili and Rathje(2009)]{saygili2009probabilistically}
Saygili, G. and Rathje, E.~M. (2009) {Probabilistically based seismic landslide
  hazard maps: an application in Southern California}.
\newblock \emph{Engineering Geology} \textbf{109}(3-4), 183--194.

\bibitem[Schmitt \emph{et~al.}(2017)Schmitt, Tanyas, Jessee, Zhu, Biegel,
  Allstadt, Jibson, Thompson, van Westen, Sato, Wald, Godt, Gorum, Xu, Rathje
  and Knudsen]{Schmitt2017}
Schmitt, R.~G., Tanyas, H., Jessee, M. A.~N., Zhu, J., Biegel, K.~M., Allstadt,
  K.~E., Jibson, R.~W., Thompson, E.~M., van Westen, C.~J., Sato, H.~P., Wald,
  D.~J., Godt, J.~W., Gorum, T., Xu, C., Rathje, E.~M. and Knudsen, K.~L.
  (2017) An open repository of earthquake-triggered ground-failure inventories.
\newblock \emph{U.S. Geological Survey Data Series 1064} .

\bibitem[Stanton and Diggle(2013)]{stanton2013}
Stanton, M.~C. and Diggle, P.~J. (2013) {Geostatistical analysis of binomial
  data: generalised linear or transformed Gaussian modelling?}
\newblock \emph{Environmetrics} \textbf{24}(3), 158--171.

\bibitem[Steger \emph{et~al.}(2020)Steger, Schmaltz and Glade]{steger2020}
Steger, S., Schmaltz, E. and Glade, T. (2020) The (f) utility to account for
  pre-failure topography in data-driven landslide susceptibility modelling.
\newblock \emph{Geomorphology} p. 107041.

\bibitem[Tanya\c{s} \emph{et~al.}(2017)Tanya\c{s}, van Westen, Allstadt,
  Nowicki, G{\"o}r{\"u}m, Jibson, Godt, Sato, Schmitt, Marc and
  Hovius]{Tanyas2017}
Tanya\c{s}, H., van Westen, C., Allstadt, K., Nowicki, A. J.~M., G{\"o}r{\"u}m,
  T., Jibson, R., Godt, J., Sato, H., Schmitt, R., Marc, O. and Hovius, N.
  (2017) Presentation and {A}nalysis of a {W}orldwide {D}atabase of
  {E}arthquake-{I}nduced {L}andslide {I}nventories.
\newblock \emph{Journal of Geophysical Research: Earth Surface}
  \textbf{122}(10), 1991--2015.

\bibitem[Tanya{\c{s}} \emph{et~al.}(2018)Tanya{\c{s}}, Allstadt and van
  Westen]{Tanyas2018}
Tanya{\c{s}}, H., Allstadt, K.~E. and van Westen, C.~J. (2018) An updated
  method for estimating landslide-event magnitude.
\newblock \emph{Earth surface processes and landforms} \textbf{43}(9),
  1836--1847.

\bibitem[Tanya{\c{s}} and Lombardo(2020)]{TanyasLombardo2020}
Tanya{\c{s}}, H. and Lombardo, L. (2020) {Completeness Index for
  Earthquake-Induced Landslide Inventories}.
\newblock \emph{Engineering Geology} p. 105331.

\bibitem[Tanyas \emph{et~al.}(2019)Tanyas, Rossi, Alvioli, van Westen and
  Marchesini]{tanyas2019}
Tanyas, H., Rossi, M., Alvioli, M., van Westen, C.~J. and Marchesini, I. (2019)
  A global slope unit-based method for the near real-time prediction of
  earthquake-induced landslides.
\newblock \emph{Geomorphology} \textbf{327}, 126--146.

\bibitem[Team \emph{et~al.}(2013)]{team2013r}
Team, R.~C. \emph{et~al.} (2013) R: A language and environment for statistical
  computing .

\bibitem[USGS(2017)]{USGS2017}
USGS (2017) {U.S. Geological Survey, 2017, USGS Earthquake Scenario Map (BSSC
  2014)}.
\newblock \emph{accessed on March 15, 2020} .

\bibitem[{Varnes and the IAEG Commission on Landslides and Other
  Mass-Movements}(1984)]{varnes1984}
{Varnes and the IAEG Commission on Landslides and Other Mass-Movements} (1984)
  Landslide hazard zonation: A review of principles and practice.
\newblock \emph{Natural Hazards, Series. Paris: United Nations Economic,
  Scientific and cultural organization. UNESCO} \textbf{3}, 63.

\bibitem[Wald \emph{et~al.}(2003)Wald, Wald, Worden and
  Goltz]{wald2003shakemap}
Wald, D., Wald, L., Worden, B. and Goltz, J. (2003) {ShakeMap, a tool for
  earthquake response}.
\newblock Technical report.

\bibitem[Wald \emph{et~al.}(1999)Wald, Quitoriano, Heaton, Kanamori, Scrivner
  and Worden]{wald1999trinet}
Wald, D.~J., Quitoriano, V., Heaton, T.~H., Kanamori, H., Scrivner, C.~W. and
  Worden, C.~B. (1999) {TriNet “ShakeMaps”: Rapid generation of peak ground
  motion and intensity maps for earthquakes in southern California}.
\newblock \emph{Earthquake Spectra} \textbf{15}(3), 537--555.

\bibitem[Worden and Wald(2016)]{worden2016shakemap}
Worden, C. and Wald, D. (2016) {ShakeMap manual online: Technical manual,
  user’s guide, and software guide}.
\newblock \emph{US Geol. Surv.} .

\bibitem[Wu \emph{et~al.}(2015)Wu, Lan, Gao, Li and Yang]{wu2015}
Wu, Y.-m., Lan, H.-x., Gao, X., Li, L.-p. and Yang, Z.-h. (2015) A simplified
  physically based coupled rainfall threshold model for triggering landslides.
\newblock \emph{Engineering Geology} \textbf{195}, 63--69.

\bibitem[Zevenbergen and Thorne(1987)]{zevenbergen1987quantitative}
Zevenbergen, L.~W. and Thorne, C.~R. (1987) Quantitative analysis of land
  surface topography.
\newblock \emph{Earth surface processes and landforms} \textbf{12}(1), 47--56.

\end{thebibliography}
\end{document}